\documentclass[aps,prb,twocolumn,showpacs,groupedaddress]{revtex4-1}  % for review and submission 
\usepackage{graphicx}  % needed for figures
\usepackage{dcolumn}   % needed for some tables
\usepackage{bm}        % for math
\usepackage{amssymb}   % for math
\usepackage{psfrag}

\begin{document}

% the following information is for internal review, please remove them for submission
%\leftline{Version xx as of \today} 
%\leftline{Primary authors: Matteo Farnesi, Joachim Reiner, Urs Sennhauser, Louis Schlapbach}
%\rightline{Comment to {\tt d0-run2eb-nnn@fnal.gov}}
%\rightline{by May 10, 2004}

% the following line is for submission 
%\hspace{5.2in} %\mbox{Fermilab-Pub-04/xxx-E}

\title{Efficient generation of realistic model systems of amorphous silica}
\author{                                                                       
%% names begin here                                                            
Matteo Farnesi Camellone,$^{1,2}$
Joachim Reiner,$^{1}$
Urs Sennhauser,$^{1}$ 
Louis Schlapbach,$^{1,2}$     
\\                                                                                                                                                                                                                                                               
}                                                                              
\affiliation{                                                                  
\centerline{$^{1}$EMPA, Swiss Federal Laboratories for Materials Testing and Research, Electronics/Metrology Laboratory,}
\centerline{8600 Duebendorf, Switzerland}  
\centerline{$^{2}$ Physics Department, Swiss federal Institute of Technology Lausanne, Ph-Ecubens, Ch-1015 Lausanne-EPFL, Switzerland}
}                                                                              
%end
\date{\today}

\begin{abstract}
We used classical molecular dynamics and the van Beest Kramer van Santen (BKS) potential to generate small model systems of amorphous silica. We further optimized the classically equilibrated configurations using plane wave based density functional theory and a generalized gradient (GGA) approximation. Within ab initio treatment we showed that both geometry optimization and Car-Parrinello annealing lead to the same final configurations but the CPU time required for the geometry optimization to reach convergence is one fifth of the time needed by a Car-Parrinello annealing. In addition during the optimization or the annealing no substantial change occurs on the topology acquired by the vitreous silica at the end of the classical quenching protocol. Structural and electronic properties have been calculated and compared to experiments.

\end{abstract}

\maketitle

\section{\label{sec:level0} INTRODUCTION}
% sections are not used for PRL papers
Disordered forms of silicon dioxide ($SiO_{2}$) glasses play a key role in electronic device applications like semiconductor devices and optical fibers. Amorphous silica has a structure consisting of a continuous random network of corner-sharing $SiO_{4}$ tetrahedra linked to a continuous three dimensional network \cite{1}. 
Numerical simulations in the framework of classical molecular dynamics (MD) has allowed in the past to clarify several issues connected with the structural properties of this material
at the microscopic level. Basis of this approach are interaction potentials fitted to {\it ab initio} and experimental parameters. 
Density Functional Theory (DFT) \cite{2} provides instead a description of both the electronic and ionic degrees of freedom and represents a powerful tool to study disordered systems. Indeed, this approach allows to improve the description of the structure and gives access to the electronic properties of amorphous $SiO_{2}$ ($a$-$SiO_{2}$) \cite{3}.
Car-Parrinello molecular dynamics (CPMD) simulation allows to study dynamical properties of the system within DFT\cite{4}. During the nuclear dynamics, the wavefunction remain sufficiently close to the Born-Oppenheimer surface allowing a faithful description of the electronic structure in most standard situation. 
However, the treatment of the dynamics of disordered systems in an ab-inito way is extremely costly from a computational point of view. This imposes restrictions to the size of the systems that can be studied and on the maximum lenght of the trajectory that can be calculated in a molecular dynamics simulation.
The lenght of the trajectory that can be obtained within the Car-Parrinello method (CPMD) \cite{4} is of the order of picoseconds. In order to overcome these limitations 
different simulation strategies have been adopted in the past that combine the advantages of classical dynamics and the accuracy of an ab initio description.
Among these, the methodology exemplified in a paper by Ginhoven et al. \cite{5} has been validated in a very accurate way.  
Through the comparison of several combinations of classical and density functional schemes, the authors show that small systems of $a$-$SiO_{2}$ (under 100 atoms) exibhit local structural characteristics that are similar to those of larger systems \cite{5}.
Benoit et al. \cite{6} use classical molecular dynamics for annealing small samples at very high temperatures (above $T=3000K$), quench the liquid up to a glassy transition, and then refine their analysis with a CPMD dynamics run followed by CPMD quenching to find a minimum energy structure.
However, the work of Benoit et al. still leaves space for improvement. In this paper we propose an enhanced protocol based on the same philosophy, but with a few important differences. First of all we use an improved classical potential, suitable for high temperature phases. Subsequently, our CPMD quenching protocol, particularly suited for the amorphous samples of small size shows to be more efficient than the one used in the past. Finally, we provide an interesting insight into the ring analysis by doing energetical considerations that should help to assess the validity of amorphous samples generated with this method.
This work shows how to efficently generate small systems that still well represent the properties of infinite sized samples.
The procedure described in this paper reduces the amount of CPU time required for the generation of $a$-$SiO_{2}$ by about a factor of five with respect to \cite{6} without compromising the final results. The properties of these system models are in good agreement with fully ab initio model systems and with available experimental data.
 
%\subsection{\label{sec:level2}Second-level heading: Formatting}
% subsections are not used for PRL papers

\section{\label{sec:level1} Computational setup}
\begin{figure}   
\includegraphics[height=70mm,width=40mm,angle=-90]{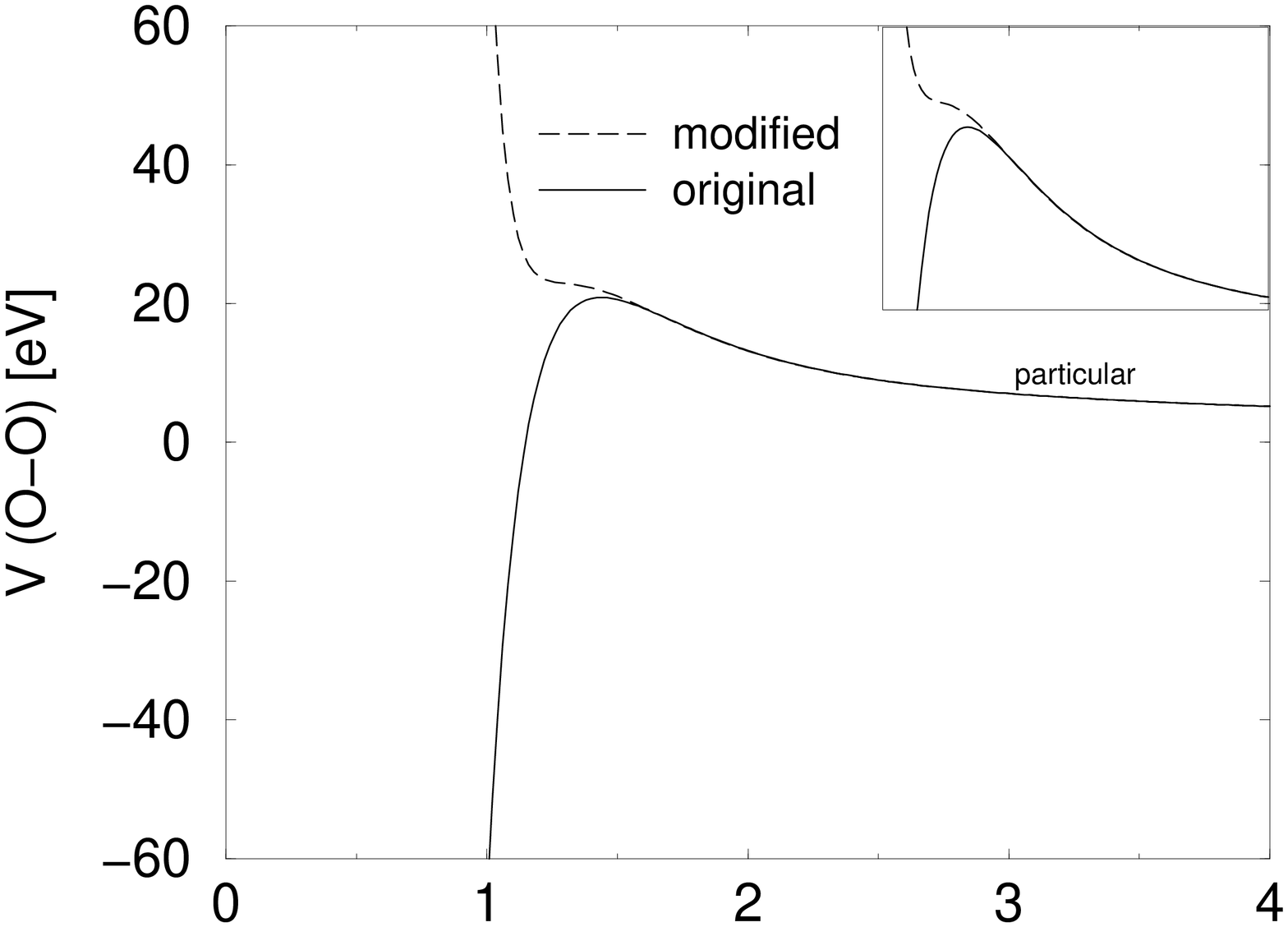}   
\hspace{5.5cm} $r_{ij}(\AA)$\\ 
\includegraphics[height=70mm,width=40mm,angle=-90]{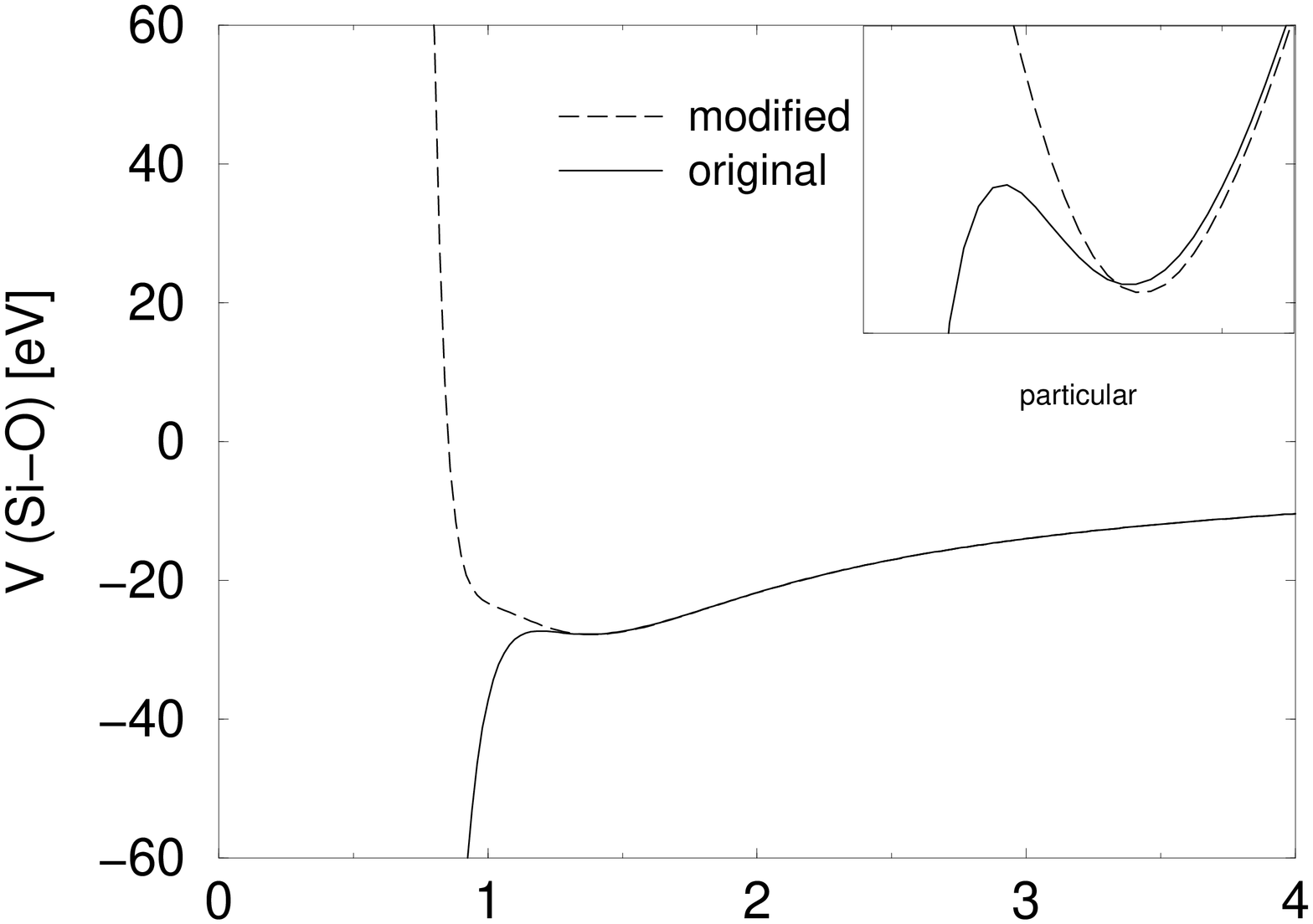}      
\hspace{5.5cm} $r_{ij}(\AA)$\\   
\includegraphics[height=70mm,width=40mm,angle=-90]{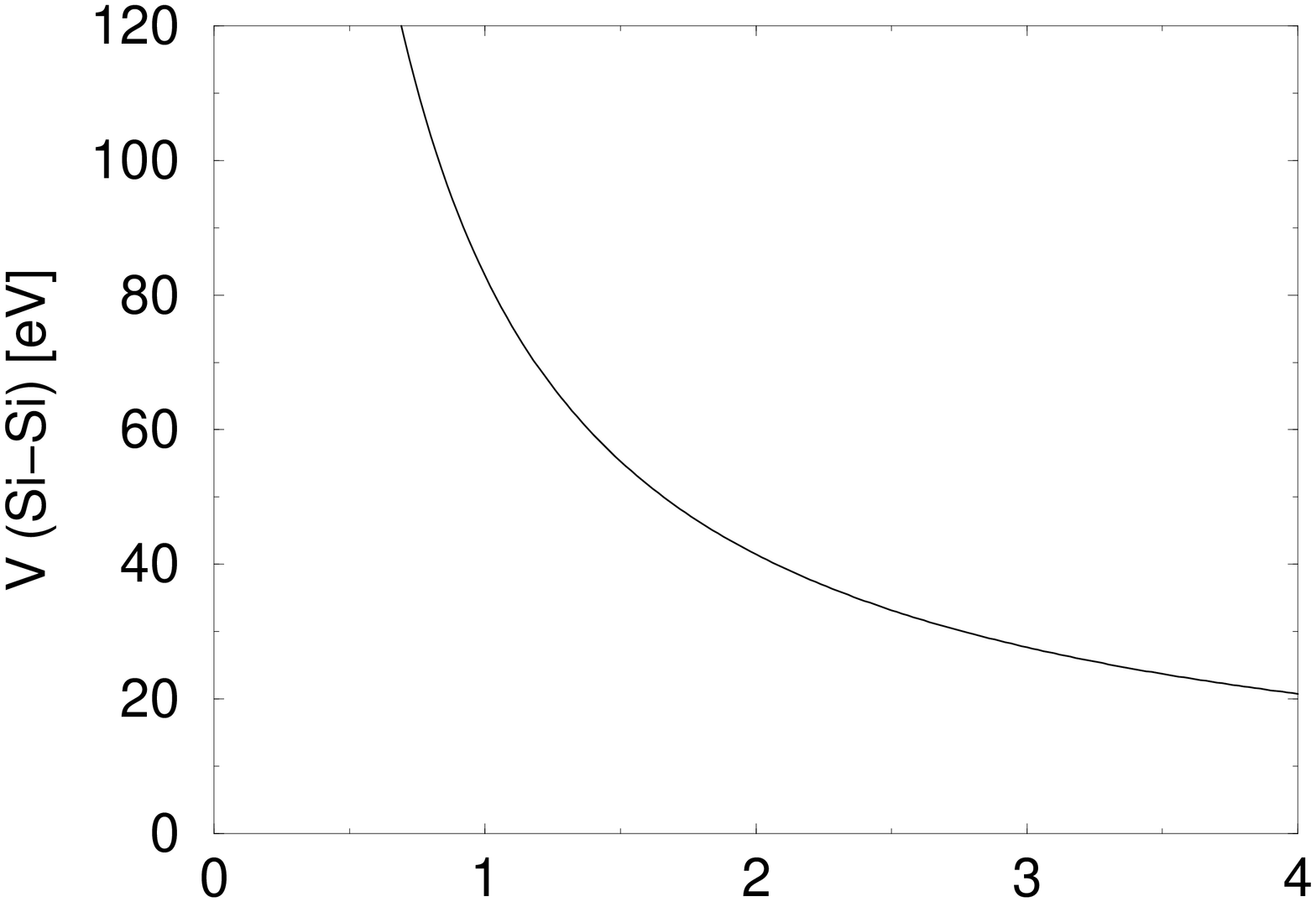}       
\hspace{5.5cm} $r_{ij}(\AA)$\\ 
\caption{Plot of the interatomic potential used in the classical MD simulation. In the upper graph we show the original BKS potential describing the interaction between O-O atoms and the modified version implemented in the simulations. In the middle graph we plot the BKS and the modified potential describing the Si-O interaction. In the lower graph we report the plot of the potential for the Si-Si interaction. In this case we have only Coulomb interactions and no short range terms.}\label{Fig. 1}  
\end{figure}

Our starting model was a slightly deformed cell of 72 atoms in the solid $\alpha$-quartz phase.
We adopted periodic boundary conditions (PBC) in all directions in order to mimic the bulk behavior. The PBC scheme allows to reduce finite size effects as well as surface effects and to get realistic properties with a small simulation cell. 
The model of the amorphous phase $a$-$SiO_{2}$ was obtained by first melting the quartz cells and then quenching from the melt using classical molecular dynamics with step cooling from $T = 8500$ to $300$ K at a quenching rate of $1.6\times10^{12}$ K/s (see section III).
We then relaxed the obtained configurations using the formalism of DFT.\\ 
The classical MD calculations were performed using the MOLDY package \cite{7}. The long range electrostatic forces were treated using the Ewald technique \cite{8}. We classically modeled the interaction between atoms using the empirical potential introduced by van Beest et al. (BKS) \cite{9} which has been shown to properly describe the structural and dynamical properties of amorphous silica. This potential is a two body interaction consisting of a Coulomb term and a short range term cast in the usual Buckingham form. Only two short range interactions are taken into account: the Si-O interaction which describes the silica bond and the O-O non bonded interaction which causes the tetrahedal arrangement of oxygen atoms around the silicon atom. At short distances the BKS potential diverges attractively limiting its use in molecular dynamics runs at high temperature (in our case above 3700 K). In order to avoid this divergence at short distances we add a repulsive term \cite{10}. This modification is necessary to describe the high temperature liquid where ions of opposite charges would otherwise approach each other very closely and be trapped in the well of the Coulomb potential.  The functional form of the potential adopted in the classical simulations is therefore: 
\begin{eqnarray}
V_{ij}=\frac{e^{2}}{4\pi\epsilon_{0}}\frac{q_{i}q_{j}}{r_{ij}}+A_{ij}e^{-B_{ij}r_{ij}}-\frac{C_{ij}}{r_{ij}^{6}} +\frac{D_{ij}}{r_{ij}^{12}}-\frac{E_{ij}}{r_{ij}^{8}}
\end{eqnarray}
where the first three terms represent the standard BKS potential and the last two terms the correction introduced to avoid the divergence at short distances. The force field parameters for the BKS potential  are reported in Table I. In Table II we report the coefficients of the correction terms applied and in Fig. \ref{Fig. 1} we plot the original and the modified BKS potential. These additional terms do not change the physics of crystalline and amorphous phases.

\begin{center}
\begin{table}
\caption{\label{table:1} Parameters of BKS potential.} 

\begin{ruledtabular}
\begin{tabular}{ccccc}
{\it i-j } & {\it $A_{ij} (eV)$} & {\it $B_{ij} (\AA^{-1})$} &  {\it $C_{ij} (eV\AA^{6})$} & {\it $charges$}  \\     \hline 
O-O & 1388.7730 & 2.76000 & 175.0000 & $q_{O}=-1.2$   \\  
Si-O & 18003.7572 & 4.87318 & 133.5381 & $q_{Si}=2.4$ \\
Si-Si & ---& --- & --- & \\

\end{tabular}
\end{ruledtabular}
\end{table}
%\end{center}

%\begin{center}
\begin{table}
\caption{\label{table:2}  Correction parameters to BKS potential.} 
\begin{ruledtabular}
\begin{tabular}{cccc} 
{\it i-j } & {\it $D_{ij} (eV \AA^{12})$} &  {\it $E_{ij} (eV\AA^{8})$} & {\it charges}   \\   \hline 
O-O &  180.00 & 24.00 & $q_{O}=-1.2$   \\  
Si-O & 20.00 & 6.00 & $q_{Si}=2.4$ \\
Si-Si & --- & --- & \\

\end{tabular}
\end{ruledtabular}
\end{table}

\end{center}

DFT calculations were performed using the CPMD \cite{11} software. The electronic structure has been treated using plane waves (PW) basis sets and pseudopotentials. The use of PW has several advantages. PW call for the use of PBC and the atomic forces can be computed via the Hellman-Feynman theorem without evaluating the Pulay contributions \cite{12}. In addition the convergence of total energy and force calculations can be controlled by a single parameter (kinetic energy cutoff) and improved to arbitrary accuracy. We made use of the Generalized Gradient approximation (GGA) \cite{13} to describe the exchange-correlation energy. Core-valence interactions have been described through norm-conserving pseudopotentials for both oxygen and silicon atoms \cite{14}. We used a plane wave basis set defined by an energy cutoff of 70 Ry for plane waves and 280 Ry for the electronic density which has been shown to be large enough to insure the convergence of the energy, in some preliminary tests on $\alpha$-quartz and $\beta$-cristobalite. The Brillouin zone has been sampled at the $\Gamma$ point only \cite{15}. All the Density Functional calculations have been performed with PBC in all three spatial directions and keeping the volume of the supercell costant.

\section{\label{sec:level2} Model systems and quenching protocol procedure}
\begin{figure}  
\includegraphics[height=70mm,width=40mm,angle=-90]{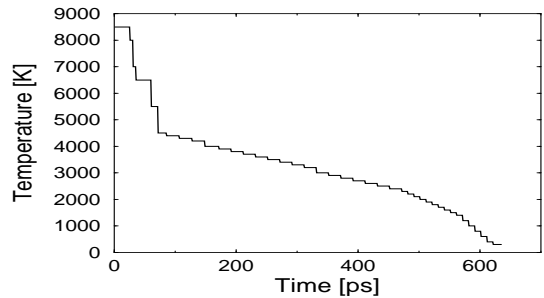}        
\caption{Quenching protocol used in this work for the generation of the amorphous SiO2 model systems.}\label{Fig. 2}   
\end{figure}

An orthorombic simulation cell of crystalline $\alpha$-quartz containing 72 atoms of Oxygen and Slicon (24 SiO2 units) has been rescaled in the $x$, $y$ and $z$ direction making it cubic and matching the amorphous $SiO_{2}$ experimental density of $2.20$ gr/cm$^{3}$. The so obtained configuration has been used as the starting supercell for the simulations. The size of this periodically repeated cubic supercell was 10.29 $\AA$. \\
In the classical MD simulations the equations of motion for the ions were integrated using a modified version of the Beeman algorithm as implemented in MOLDY \cite{7} with a timestep of 1.6 fs. In total seven amorphous silica model systems were generated. During the simulations all the atoms within the supercell have been allowed to move while the volume of the supercell has been kept fixed. The starting configurations have been heated up rapidly to $T=8500$ K leading to the formation of liquid silica. We subsequently performed a 25 ps costant temperature run at $T=8500$ K in order to equilibrate these high temperature liquids. We then quenched the model systems to $T=300$ K using constant temperature MD runs performed at different temperatures, with a final averaged quenching rate of  $1.6\times10^{12}$ K/s. The quenching protocol is reported in Fig. \ref{Fig. 2}. It is important to notice that the quenching rate should be not too high. Very high quenching rates hinder an adequate relaxation, and can eventually freeze the concentrations of small rings (see section IV) to values much higher than those experimentally expected. During all the costant temperature runs the velocities of the particles were periodically rescaled assuming a Maxwell-Boltzmann distribution at the target temperature. We noticed from the mean square diplacements of the Si and O atoms, that the freezing temperature of these model systems can be located around $T=3300$ K, which is much larger than the experimental data. The resulting structures were perfectly ordered with every silicon atom fourfold and every oxygen atom twofold coordinated. The final configurations were equilibrated classically at $T=300$ K for about one nanosecond. We finally performed a 0.5 picosecond microcanonical Car-Parrinello molecular dynamics run on the equilibrated systems ($T=300$ K) in order to get configurations for structural analysis and comparison with the classical model systems. The analysis of the structural properties is reported in section VI. 

\section{\label{sec:level3} AB INITIO STRUCTURAL RELAXATION}
\begin{figure}    
\centering   
\includegraphics[height=75mm,width=70mm,angle=-90]{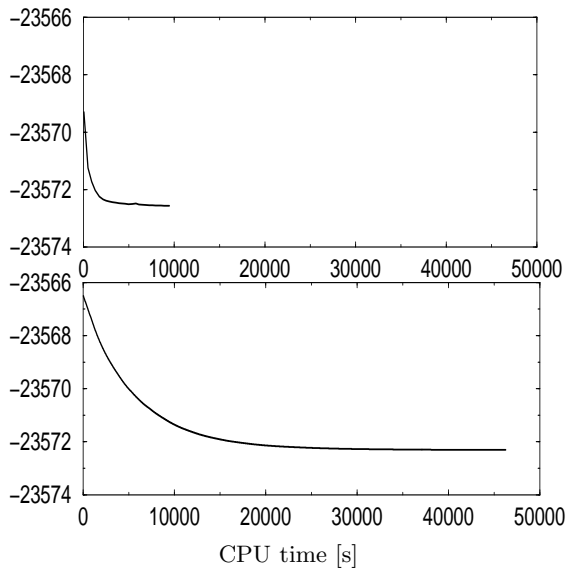}    
\vspace{2mm}  
\hspace{7cm} CPU time [s]\\     
\caption{Total energy of the system versus the CPU time in seconds with four processors. The upper graph shows the behaviour of the system in a geometry optimization, and the lower graph shows the behaviour in a Car-Parrinello simulated annealing.}\label{Fig.5}   
\end{figure}

The configurations obtained as described in the previous section have been further optimized by explicitly treating the electronic degrees of freedom with DFT. DFT combined with Molecular Dynamics offers several methods for structure relaxation and global optimization. In this section we describe the effects of geometry optimization (G.O.) and Car-Parrinello simulated annealing (CP annealing) of the model systems. G.O. and CP annealing represent two different ways to reach the equilibrium configuration of a given system. In a G.O. one starts with a given structure characterized by the nuclear coordinates. By evaluating the forces for that set of nuclear coordinates one gains information on how to shift the atomic positions in order to reduce the total energy. The procedure is repeated until the structure does not change any further within a predefined tolerance. An alternative approach is the CP annealing. Within this method the electronic and nuclear degrees of freedom are optimized simultaneously, starting the calculation at a given temperature. Then the temperature is reduced gradually and the structure should get trapped in that of the global energy minimum with a higher probability and not in a local one \cite{16}. This method has been shown to be successful for large systems where the energy surface is extremely frustrated.\\
When switching from the Classical MD treatment to the ab-initio description, a common way to treat the system is to first perfom a sufficentely long constant temperature (NVT) run, say at $T=300$ K  for equilibration, and then anneal to a $T=0$ K configuration to reach the ground state energy of the system \cite{6,17}. This procedure costs a lot of CPU time (in \cite{6,17} larger systems have been studied). We noticed that, at least for these small systems, a G.O. and a CP annealing lead to the same final configuration, but the CPU time required for the G.O. to reach convergence is one fifth of the time needed by a CP annealing (see Fig. \ref{Fig.5}). In addition during the optimization or the annealing no substantial change occurs on the topology acquired at the end of the classical quenching protocol. The root mean square displacement (RMSD) between the final configurations obtained within G.O. and CP annealing methods has been evaluated to be on average 0.012 $\AA$ while the mean difference in energy is 0.002 eV. The same convergence criteria ($10^{-7}$a.u. for the energy and $10^{-5}$ a.u.$\cdot\AA^{-1}$ on the gradients) have been adopted in both procedures. In the G.O. we used the limited-memory BFGS \cite{18} method as implemented in CPMD. In the CP annealing the parameters that can be controlled are the fictitious or inertia mass $\mu$, the scaling factors for ions and electrons respectevely $\gamma_{N}$ and $\gamma_{e}$ and the time step \cite{19,20,21}. We choose $\mu$=700 a.u., $\gamma_{N}$=0.99, $\gamma_{e}$=0.99 and a time step of 0.12 fs. We further investigated the relaxation process by applying a combination of CP annealing and G.O. to the model systems. In table \ref{tab:3} we report the CPU time needed to reach the convergence by applyng first a CP annealing for a certain number of time steps followed by a G.O. As can be inferred from the table the CPU time is minimum when no annealing is performed and the structure is simply geometry optimized.

\begin{figure}     
\centering    
\includegraphics[scale=0.3,angle=-90]{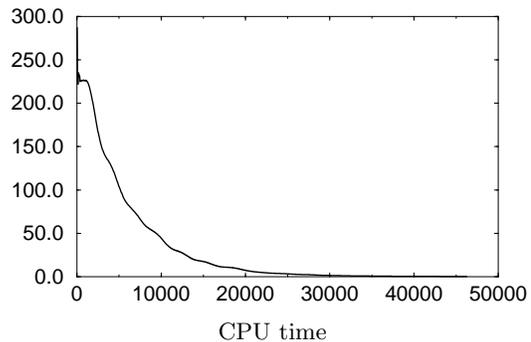}     
\vspace{2mm}   
\hspace{7cm} CPU time\\      
\caption{Temperature behavour of the system during the same Car-Parrinello simulated annealing represented in the lower graph of Fig. \ref{Fig.5}.}\label{Fig.6}    
\end{figure}

\begin{center}
\begin{table}
\caption{\label{tab:3} CPU time in seconds with four processors needed to relax the system to its ground state as a function of the number of time steps of the initial CP run.}
\begin{ruledtabular} 
\begin{tabular}{cccccccc}  
{\it Step } & {\it 0} & {\it 100} & {\it 200} & {\it 300} & {\it 500} & {\it 700} & {\it 2000}   \\   \hline  
CP (s) &  0 & 2314 & 4628 & 6943 & 11572 & 16201 & 46289 \\   
G.O. (s) & 9731 & 9340 & 10059 & 9602& 6819 & 5779 & 0 \\ 
Total (s) & 9731 & 11654 & 14687 & 16545& 18391 & 21980 & 46289\\ 
\end{tabular} 
\end{ruledtabular}
\end{table} 
\end{center}

\section{\label{sec:level4} Criteria for selecting the model systems and ring analysis} 
One way to characterize $SiO_{2}$ networks is by studying the statistics of closed rings of bonded atoms \cite{22}. An $N$ ring is described as a closed $Si$-$O$-$Si$-$O$$\cdot\cdot\cdot$ chain with $N$ silicon atoms \cite{23}. Small rings ($N$=2 or 3) are geometrically strained structures and result in an energy penalty \cite{24}.\\
During the cooling protocol unrealistic geometry structures can occur. Two reasons for the appearence of such unphysical structures are the limited size of the computational box (72 atoms) and the applied PBC. In addition a too high quenching rate can prevent an appropriate relaxation of the model systems and freezes the concentration of small rings \cite{3}. Unrealistic model systems exhibit the presence of large voids and compressed regions within the simulation box or concentrations of small membered-rings much higher than the experimental ones \cite{3}. While large voids can be consistent with the structure of amorphous silica, the high density of compressed regions of the cell is not. For these reasons we generated several model systems of $a$-$SiO_{2}$ and selected the ones which satisfied all the criteria of a realistic structure. A primary criterion for selecting the systems was to compute the total energy and eliminate the model systems with higher energies corresponding to higher stresses. As a consequence of this criterion systems which exibihited a too high concentration of small rings, which are strained structures, have been eliminated.  Among seven generated systems only one was eliminated because it showed a too high concentration of small rings confirming that the quenching protocol adopted was slow enough for an appropiate relaxation of the amorphous configurations.\\ 
The three-membered rings are quasiplanar as can be deduced from the sum over all bond angles in the rings that average to $702^{\circ}$, while the ideal value is $720^{\circ}$. The average Si-O-Si angle in these rings is $131.80^{\circ}$ which is smaller than the average of $147^{\circ}$ of the model systems. The average Si-O bond lenghts is 1.642$\AA$ in agreement with \cite{24}. Four-membered rings do not exhibit preferences for planarity; the sum over all bond angles in the rings give an angle of $976.88^{\circ}$ (ideal $1080^{\circ}$) while the average Si-O-Si angle is $137.02^{\circ}$, which is smaller than the reference angles inferred from NMR measurements ($145^{\circ}-150^{\circ}$ \cite{ANGLE}) or from x-ray diffraction experiments ($144^{\circ}$ \cite{26}).\\
In table \ref{tab:4} we report the differences $\Delta E$ of the total energy of the model systems to the one with the lowest energy, which we called model-4. In the other columns we give a description of the systems in terms of ring statistics. As can be seen from the table the systems with higher energies and stress are that ones with a higher concentration of small rings, showing that the presence of small rings is the main cause of stress in the system. We estimated an upper bound fo $\Delta E$ of 1 eV per small rings (2,3 membered rings) which is in agreement with \cite{24}, considering the fact that in additon to the penalty energies coming from the strained rings other sources for stress can be the fact that during all the simulations the volume of the computational box has been kept fixed.

\begin{center}
\begin{table}
\caption{\label{tab:4} In the second column $\Delta E$ is the difference of the total energy of the model systems to the one with the lowest energy. In the other columns number of n-rings in the model systems.}

\begin{ruledtabular} 
\begin{tabular}{ccccccccc} 
{\it models } & {\it $\Delta E$(eV)}  & {\it 2-R} &  {\it 3-R} & {\it 4-R} & {\it 5-R} & {\it 6-R} & {\it 7-R} & {\it 8-R} \\   \hline  
model-1 &1.15  & 0 &1  &1 &1 &1 &1 &0  \\   
model-2 & 1.65 & 0 &3  &1 &2 &0 &1 &1  \\ 
model-3 & 0.7 & 0 & 0 & 1&1 &3 &2 &1 \\   
model-4 & 0 & 0 & 0 & 1 & 1 & 2 & 1 & 0  \\ 
model-5 &0.44 & 0 & 0 & 2 & 1 &2 & 1 & 1 \\
model-6 &1 & 0 & 2 & 1 & 3 & 2 & 1 & 0 \\
model-7&1.2 & 1 & 2 & 1 & 4 & 3 & 1 & 0\\

\end{tabular} 
\end{ruledtabular}
\end{table} 
\end{center}

\section{\label{sec:level5} Structural properties}
\begin{figure}   
\centering  
\includegraphics[height=65mm,width=65mm,angle=-90]{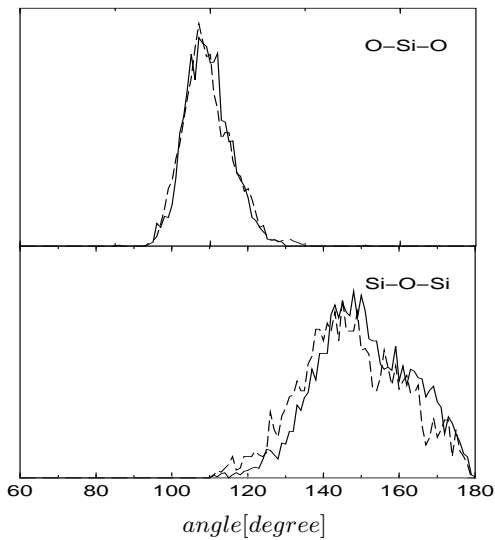}   
\vspace{2mm} 
\hspace{6cm} $angle[degree]$\\    
\caption{Averaged angle distribution for O-Si-O (upper graph and Si-O-Si (lower graph). Solid lines refer to Car-Parrinello MD and dashed lines refer to Classical MD simulations.}\label{Fig. 4}  
\end{figure}

\begin{figure}  
\centering 
\includegraphics[height=70mm,width=85mm,angle=-90]{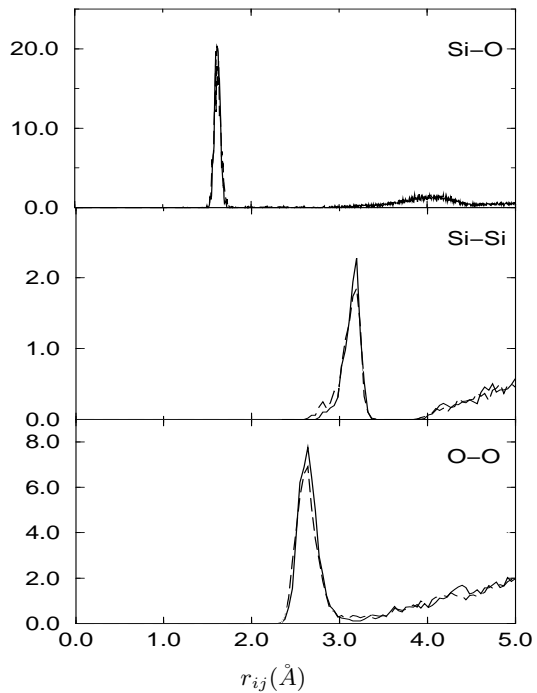}  
\vspace{2mm} 
\hspace{6cm} $r_{ij}(\AA)$\\   
\caption{Simulated pair correlation functions for Si-O (upper graph), Si-Si (middle graph) and O-O (lower graph). The pair correlation functions are obtained from NVE simulations at 300 K. The solid lines refer to Car-Parrinello runs while the dashed lines refer to Classical MD runs.}\label{Fig. 3} 
\end{figure}

Among all the generated $SiO_{2}$ model systems we selected the model-4 which best satisfied our criteria for a detailed study.  
This choice has been performed on the basis of the ring analysis of the previous section, but further analysis has to be performed to validate
this finite size sample.
The most natural way to further estimate the quality of the generated amorphous model systems is to analyze their structural characteristics such as bond lenghts and angle distributions, radial distribution functions, static structure factors and to compare them to available experimental data. We evaluated all these properties for the system called model-4. At $T=300$ K we have calculated the time-averaged distributions of the intra-tetrahedral O-Si-O and inter-tetrahedral Si-O-Si angles for the classical and quantum MD simulations using the same starting configuration (Fig. \ref{Fig. 4}). The intra-tetrahedral angles O-Si-O stay close to the experimental value of 109.4$^{\circ}$ for both simulations: 109.3$^{\circ}$ $\pm$ 5 for the classical and 109.5$^{\circ}$ $\pm$ 7 for the CPMD. The only difference between the classical and quantum simulation of the angle distibutions is a small shift of the mean value of the intra-tetrahedral Si-O-Si angle. We estimated for this angle a mean value of 152$^{\circ}$ $\pm$ 11$^{\circ}$ for the classical averaged configurations and 146$^{\circ}$ $\pm$ 6$^{\circ}$ for the Car-Parrinello microcanonical run (NVE) (the experimental value being 140$^{\circ}$-150$^{\circ}$\cite{25}). We noticed that this change in angle takes place during the first steps of the Car-Parrinello run corresponding to an extremely short sampling time ($\approx$ 0.07 ps).\\ 
We have then calculated the averaged pair correlation functions $g_{\alpha\beta}(r)$, where $\alpha$, $\beta$= Si, O of the systems at $T=300$ K for the Classical and Car-Parrinello runs (Fig. \ref{Fig. 3}). Within the classical treatment we obtained an averaged distance of 1.61$\AA$ for Si-O, 3.12$\AA$ for Si-Si, and 2.66$\AA$ for O-O, in very good agreement with the experimental data \cite{27}. For the Car-Parrinello treatment the averaged bond lenght Si-O is found to be 1.65$\AA$, which is slightly (3$\%$) larger than the experimental value (1.61$\AA$), in agreement with a general tendency of the GGA \cite{28}. Using the same computational setup we found a similar overstimation in a preliminary test made on $\alpha$-quartz. For the other peaks we found 3.18$\AA$ for Si-Si and 2.68$\AA$ for the O-O bond lenght.   \\
From this analysis we concluded that our classical systems were well equilibrated. 
From the pair correlation functions we computed the static structure factor for the Car-Parrinello run which can be compared with experiments \cite{29}. Experimentally the static structure factor can be obtained from neutron diffraction. In the simulations one possibility is to compute the static structure factor $S(q)$  by its relation  to the pair correlation function $g(r)$:

\begin{eqnarray}
S(q)=1+4\pi\rho\int_{0}^{\infty}(g(r)-1)\frac{sin(qr)}{qr}r^{2}dr \label{sq}
\end{eqnarray}

with

\begin{eqnarray}
g(r)=\frac{\Bigg[\sum_{\alpha,\beta}c_{\alpha}b_{\alpha}c_{\beta}b_{\beta}g_{\alpha\beta}(r)\Bigg]}{\Bigg[\sum_{\alpha}c_{\alpha}b_{\alpha}\Bigg]^{2}} \label{matto}
\end{eqnarray}
The integral in Eq.(\ref{sq}) has been evaluated by using the Filon's method \cite{30}. In equation (\ref{sq}) $\rho$ stands for the density. In Eq. (\ref{matto}) the $g_{\alpha\beta}(r)$ are the pair correlation functions ($\alpha, \beta$ = Si,O) while $c_{\alpha}$ and $c_{\beta}$ stand for the concentrations of the two species and $b_{\alpha}$ and $b_{\beta}$ are their scattering lenghts. We chose $b_{\alpha}$ and $b_{\beta}$ to be 4.149 fm for Si and 5.803 fm for O \cite{31}. The integration from 0 to $\infty$ is performed as a summation from zero to half of the box size. The first sharp diffraction peak (FSDP) corresponds to 1.53$\AA^{-1}$ which is in good agreement with the eperimental data (1.52$\AA^{-1}$) \cite{31}. The other peaks are at 2.8, 5.19, and 7.83$\AA^{-1}$, respectevely.\\
We finally studied the electronic density of states of the system. We evaluated the Kohn-Sham density of states (Fig. \ref{Fig. 10}). The density is in agreement with other DFT calculations \cite{32,6}. The calculated band gap is 5.7 eV and it is understimated (experimental $\sim$ 9 eV) as usual in DFT due to the discontinuity of derivatives of the exchange-correlation functional \cite{33}. The DOS of amorphous and crystalline $SiO_{2}$ are very similar, because the short range order is similar for both structures. For this reason we can identify the states at about -20 eV as oxygen $2s$ states, the states from -10 eV to -4 eV as bonding states between silicon $sp^{3}$ hybrids and oxygen $2p$ orbitals and above -4 eV as oxygen $2p$ nonbonding orbitals. The lowest conduction band are states with antibonding character \cite{34}. 
\\

\begin{figure} 
\includegraphics[height=67mm,width=37mm,angle=-90]{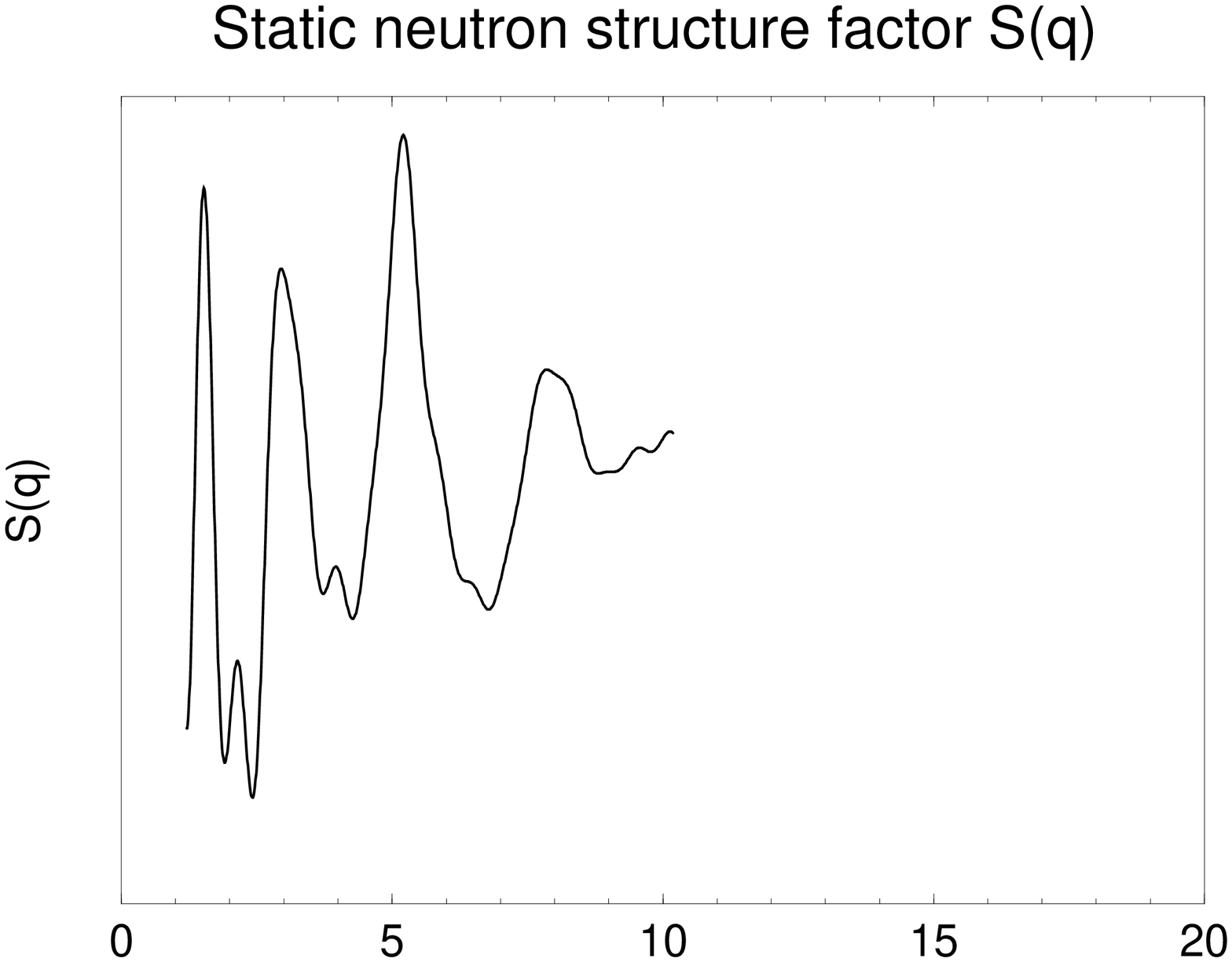}  
\includegraphics[scale=0.338]{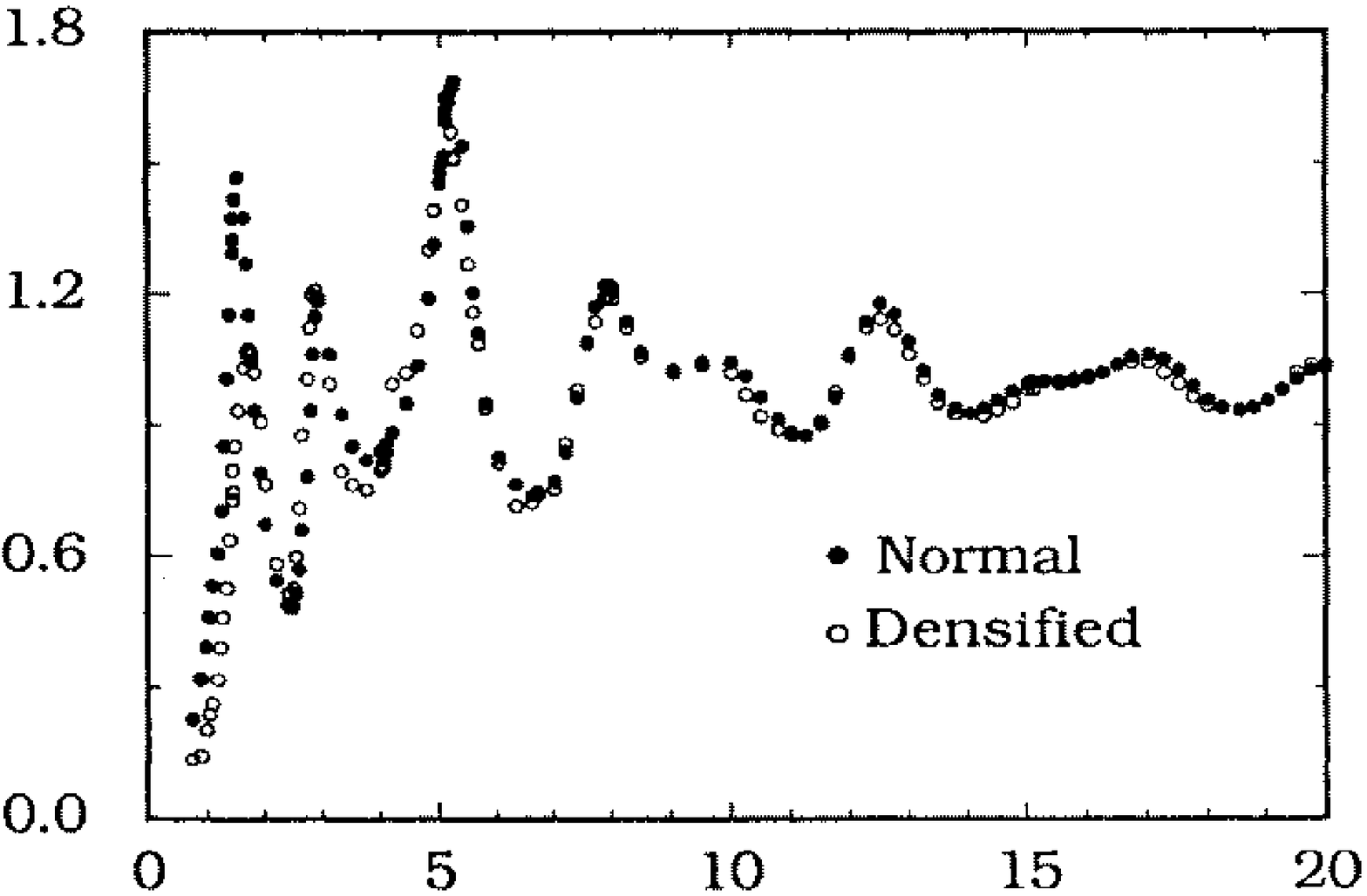}
\vspace{2mm}   
\hspace{7cm} q ($\AA^{-1})$\\    
\caption{\label{fig:3} S(q) at 0 K computed with CPMD (upper graph), and experimental results from neutron diffraction of normal and densified vitreous $SiO_{2}$ (from Ref. \cite{31}) (lower graph).}  
\end{figure}

\begin{figure} 
\includegraphics[height=72mm,width=42mm,angle=-90]{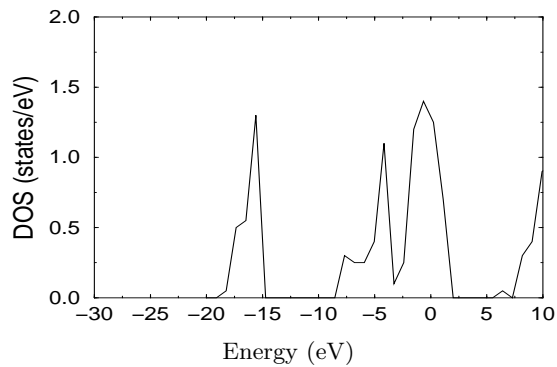}    
\vspace{2mm}   
\hspace{7cm} Energy (eV)\\
\caption{Electronic density of states of amorphous SiO2 at 0K.}\label{Fig. 10} 
\end{figure}

\section{\label{sec:level6} Conclusions}
Using classical molecular dynamics and the BKS empirical potential we generated seven model systems of 72 atoms of $a$-$SiO_{2}$ at 300K. We further optimized the classical equilibrated configurations with plane wave based density functional theory and GGA approximation to the exchange correlation energy. In particular within ab initio treatment we showed that both geometry optimization and Car-Parrinello annealing applied to our systems lead to the same final configurations but the CPU time required for the geometry optimization to reach convergence is one fifth of the time needed by a Car-Parrinello annealing. The topology acquired by the amorphous silica model systems at the end of the classical quenching protocol did not change when treated in an ab initio way. The structural properties of the amorphous glass, like angle distributions, pair correlation functions, static structure factor and electronic density of states have been accurately studied. The properties of rings in model systems have been accurately analyzed by energetical considerations. We found very similar results for both the description (classical and Car-Parrinello) and in good agreement with experimental data.\\
The generated model systems will be used to study defects, and in particular the mechanism for bond weakening and electron localization in $SiO_{2}$.

\section*{ACKNOWLEDGEMENTS}
Two of the authors, M. Farnesi Camellone and J. Reiner, would like to thank J. Hutter, Ari P. Seitsonen, D.Donadio for helpful discussions. M. Farnesi Camellone thanks Daniele Passerone for carefull reading of the manuscript. This work was supported by the Swiss Nanotechnology Program TOP NANO 21 (CTI No:TNS-NM 6499.1). All the calculations were performed on IBM-SP4 at the Swiss National Computing Center (CSCS).


\begin{thebibliography}{99}
  
  \bibitem{1} 
    {\sl The Physics and Technology of Amorphous $SiO_{2}$}, edited by R.A.B. Devine (Plenum Press, New York, 1988).

   

  \bibitem{2} 
    P.Hohenberg and W.Kohn, Phys. Rev. {\bf 136} (1964), 
    W.Kohn and L.J.Sham, Phys. Rev. {\bf 14} (1965).

  \bibitem{3}
    J.Sarnthein, A.Pasquarello and R.Car, Phys. Rev. B {\bf 52}, 12690 (1995).


  \bibitem{4}
    R.Car amd M.Parrinello, Phys. Rev. Lett. {\bf 55}, 2471 (1985).
  
  


  \bibitem{5}
    R.M.van Ginhoven, H. Jonsson and L.R. Corrales Phys. Rev. B {\bf 71}, 24208 (2005).

   \bibitem{6}
     M.Benoit, S.Ispas, P.Jund and R.Jullien, Eur. Phys. J. B {\bf 13}, 631, (2000).

  \bibitem{7}
    K. Refson, Comput. Phys. Comm. {\bf 126}, 310 (2000).

  \bibitem{8}
    M.P.Allen and D.J.Tildesley, {\sl Computer simulations of liquids}, Clarendon Press, Oxford, 1987, p.156.


  \bibitem{9}
    B. van Beest, G. Kramer, and R. van Santen, Phys. Rev. Lett. {\bf 64}, 1955 (1990).

  \bibitem{10}
   The correction we apported is not the only one possible see for example Y.Guissani and B.Guillot J. Chem. Phys. {\bf 104} (19), 7633,    (1996).



  \bibitem{11} 
    CPMD code by J. Hutter {\sl et al.}, Max-Planck-Insitut FKF and IBM Zurich Reasearch Laboratory, 1995-2005.

  \bibitem{12}
   P.Pulay, in {\sl Ab Initio Methods in Quantum Chemistry II}, edited by K.P.Lawley (Wiley, Chichester, 1987) p. 241. 

  \bibitem{13}
    J.P.Perdew et al. Phys. Rev. B {\bf 46}, 6671 (1992).

  \bibitem{14}
    N.Troullier, J.L.Martins, Phys. Rev. B {\bf 43}, 1993 (1991).    
  
  \bibitem{15} G.Galli and A.Pasquarello, {\sl M.P.Allen and D.J.Tildesley (eds.), Computar Simulation in Chemical Physics}, 261-313 (1993) Kluwer Academic Publishers. printed in Netherlands. 



  \bibitem{16} Michael Springborg, {\sl Methods of Electronic-Structure Calculations From Molecules to Solids}, Wiley, 2000.

  \bibitem{17}
  D.Donadio, M.Bernasconi, F.Tassone, Phys. Rev. B {\bf 68}, 134202, (2003)



 
  \bibitem{18}
  S.R.Billeter, A.Curioni and W. Andreoni, Comput. Mat. Sci (27), 437 (2003).



   \bibitem{19}
  G.Pastore, E.Smargiassi and F.Buda, Phys. Rev. A {\bf 44}, 6334 (1991).
  
  \bibitem{20}
  D.Marx and J.Hutter {\sl Modern methods and Algorithms of Quantum Chemistry}, J.Grotendorst (Ed.), J. von Neumann Institute for Computing.,  Julich, NIC series, Vol.1, pp.301  -449, 2000.
  
  \bibitem{21}
  F.Tassone, F.Mauri and R.Car, Phys. Rev. B {\bf 50}, 10561 (1994).



  \bibitem{22}
   Crystal quartz is characterized by six and eight membered rings, see for example A.C.Wrigth and J.A.E.Desa, Phys. Chem. Glasses {\bf 19}, 140, (1978).

 

  \bibitem{23}
   S.V.King, Nature {\bf 213}, 1112 (1967). 
  
  
  \bibitem{24}
    D.R.Hamann Phys. Rev. B {\bf 55} 14784 (1997).
  

  \bibitem{25}
    R.Dupree, R.F.Pettifer, Nature {\bf 308}, 523 (1991).

  \bibitem{26} R.Mozzi and B.Warren, J. App. Crystallogr. {\bf 2}, 164, (1969). 
 
  \bibitem{27}
   P.A.V. Johnson, A.C. Wright, R.N.Sinclar, J.Non-Cryst.Solids {\bf 58}, 109, (1983).
   
  \bibitem{28}
   A.Dal Corso et al. Phys. Rev. B {\bf 53}, 1180 (1996).

  
   \bibitem{29}
   S.Susman et al., Phys. Rev. B {\bf 43}, 1194 (1991).


   \bibitem{30}
   Filon, Proc. Poy. Soc. Edin., {\bf 4938}, (1928).
  

  \bibitem{31}
   S.Susman, K.J. Volin, D.L.Price, M.Grimsditch, J.P.Rino, R.K.Kalia, P.Vashishta, G.Gwanmesia, Y.Wang, R.C.Liebermann, Phys.Rev.B {\bf    43},1194,(1991).

  \bibitem{32}
  J.Sarnthein, A.Pasquarello and R.Car Phys. Rev. Lett. {\bf 74}, 4682 (1995).

  
  \bibitem{33}
   J.P.Perdew and M.Levy, Phys. Rev. Lett. {\bf 51}, 1884, (1983). 


  \bibitem{34}
   N.Bingelli, N.Troullier, J.L.Martin and J.R.Chelikowsky, Phys. Rev. B {\bf 44}, 4471 (1991).   
  


\end{thebibliography}
\end{document}